\documentclass[aps,onecolumn]{revtex4}
\usepackage{amsmath}
\usepackage{amssymb,epsf}
\usepackage{xcolor}
\usepackage{soul}
\usepackage{float}
\usepackage{graphicx}
\usepackage[caption=false]{subfig}
\begin{document}

\title{Metamaterials mimic the black holes:\\
the effects of charge and rotation on the optical properties}
\author{S. H. Hendi$^{1,2}$\footnote{email address: hendi@shirazu.ac.ir},
Z. S. Taghadomi$^{1}$ \footnote{email address:
z\_s\_taghadomi@yahoo.com} and A. Ghasempour Ardakani$^{1}$\footnote{email
address: aghasempour@shirazu.ac.ir} } \affiliation{$^1$Physics
Department and Biruni Observatory, College of Sciences, Shiraz
University, Shiraz 71454, Iran\\
$^2$ Research Institute for Astronomy and Astrophysics of Maragha
(RIAAM), P.O. Box 55134-441, Maragha, Iran}

\begin{abstract}
Motivated by investigation of black hole properties in the lab,
some interesting subjects such as analogue gravity and
transformation optics are generated. In this paper, we look for
the analogies between the geometry of a gravitating system and the
optical medium. In addition, we recognize that appropriate
metamaterials can be used to mimic the propagation of light in the
curved spacetimes and behave like black holes. The resemblance of
metamaterials with Kerr and Reissner-Nordstr\"{o}m spacetimes is
studied. At last, we compare the full-wave numerical calculation
of light with its optical limit of geometry.

\end{abstract}

%\keywords{wcwececwc ; wecwcecwc}

\maketitle

\section{Introduction}
One of the interesting subjects in gravitational physics is
analogue gravity in the optical frameworks which is attributed to
the pioneer work of Gordon \cite{Gordon1923}. Some of the Gordon's
activities are devoted to describing dielectric media by an
effective metric or strictly speaking, mimicking a gravitational
field with a dielectric medium. With the advent of transformation
optics and analogue gravity, one can create systems which closely
resemble to general relativity objects such as black holes.
Regarding the analogue gravity, one can investigate the
resemblances of general relativistic gravitational fields with the
equation of motion of other physical systems
\cite{analoguegravity}, such as mimicking some aspects of black
holes with super fluids in the Bose-Einstein condensation
\cite{becondenstate}. Although such systems may not have exact
analogues of any real spacetimes with corresponding metric
tensors, transformation optics, where an optical material appears
to perform a transformation of space, may use a formalism of
general relativity to build gradient index static media in order
to control the light paths in these media
\cite{leonhardt1,leonhardt2}. Theoretical concept of
transformation optics is similar to equations of general
relativity that describe how gravity warps space and time.
However, instead of space and time, these equations show how light
can be directed in a chosen manner, analogous to warp space
\cite{leonhardtbook}. In addition, transformation optics applies
metamaterials to produce spatial variations, derived from the
coordinate transformation. Therefore, it is possible to construct
new artificial composite devices with a desired permittivity and
permeability. Moreover, interactions of light and matter with
spacetime, as predicted by the general relativity, can be studied
by using a new type of artificial optical materials that feature
extraordinary abilities to bend light. This research creates an
interesting link between the newly emerging field of artificial
optical metamaterials and that of general relativity, and
therefore, it opens a new possibility to investigate various
general relativity phenomena in a laboratory setting: chaotic
motions observed in celestial objects that have been subjected to
gravitational fields \cite{celedtialmechanics,leonhard3},
reproduce an accurate laboratory model of the physical multiverse
\cite{leonhard3, multiverse}, optical analogues of black holes
\cite{obh1, obh2, obh3, obh4, obh5}, Schwarzschild spacetime
\cite{schwarz1, schwarz2, schwarz3} and Hawking radiation
\cite{hawking}.

The main purpose of this paper is to generate an optical analogue
of the known gravitational black hole and its characteristic
phenomena, such as the existence of an event horizon (a one-way
membrane) and the bending of light rays. Optical materials can
mimic the geometry of light near black holes, both theoretically
and experimentally \cite{leonhardtbook,obh1,obh5}. It has been
shown that the permittivity and permeability tensors can be used
instead of refractive index in order to build the equivalent
optical medium mimicking black holes
\cite{leonhardtbook,schwarz3}. In Ref. \cite{schwarz3} the
propagation of light waves outside the Schwarzschild black hole
were simulated and the results were compared with ray paths
obtained from the Hamiltonian method. Interestingly, the
phenomenon of "photon sphere" which is an important feature of the
black hole systems, was observed.

Here as a first step, we generalize the results of Ref.
\cite{schwarz3} to the case of charged black hole. We consider a
charged black hole and its optical simulation, and discuss the
effect of electric charge on the photon sphere. Motivated by the
lacking of a definitive answer on the existence of astrophysical
non-rotating black holes, we are encouraged to look at Kerr
spacetime and look for its closely optical simulation as the
second step. In addition, we were interested in the effect of spin
parameter on the photon sphere. The basic structure of this paper
is twofold. First, we determine the permittivity and permeability
mimicking the desired spacetime. Then, the Maxwell equations are
numerically solved by using constitutive relations. Second, the
Lagrangian for each spacetime is obtained. Then, by using the
equations of motion, one can find the ray paths in each spacetime.
Finally, in order to check the accuracy of simulations, we compare
the propagation of waves in optical materials with the ray paths
driven from the equations of motion.

\section{Spacetime geometry and media}

Using the mathematical machinery of differential geometry, one can
write Maxwell equations in arbitrary coordinates and geometry. It
has been shown that the source-free Maxwell equations in arbitrary
right-handed spacetime coordinates can be written as the
macroscopic Maxwell equations in the right-handed Cartesian
coordinates with Plebanski's constitutive equations
\cite{leonhardtbook}. It means that Maxwell equations in curved
coordinates are equivalent to their standard form for the flat
space but in the presence of an effective medium. The Plebanski's
constitutive relations have been found in the form
\cite{plebanski}:
\begin{equation}
\begin{array}{c}
D^i=\epsilon_0 \epsilon^{ij}E_j+\frac{1}{c}[ijk]\Gamma_j H_k,\hspace{0.5cm} B^i=\mu_0 \mu^{ij}H_j-\frac{1}{c}[ijk]\Gamma_j E_k,
\end{array}\label{eq:plebanski}
\end{equation}
where permittivity and permeability tensors and a vector $\boldsymbol{\Gamma}$ are  given by:
\begin{eqnarray}
\epsilon^{ij}=\mu^{ij}=-\frac{\sqrt{-g}}{g_{00}}g^{ij},\hspace{0.5cm} \Gamma_i=\frac{g_{0i}}{g_{00}} \label{eq:epsilonmu}.
\end{eqnarray}
Here, $g^{ij}$ is the inverse of $g_{ij}$, the spatial part of
full spacetime metric $g_{\mu \nu}$, and $g$ is the determinant of
$g_{\mu \nu}$. Designing artificial materials with the
permittivity and permeability introduced in Eq.
(\ref{eq:epsilonmu}) enables us to mimic different spacetimes. It
is notable that all the information about the gravitational field
is embedded in the material properties of the effective medium.

Anisotropic or isotropic materials in which the constitutive
relations described by Eq. (\ref{eq:plebanski}) are called
bianisotropic or bi-isotropic materials \cite{bianisotropic1,
bianisotropic2,bianisotropic3,bianisotropic4,bianisotropic5}. The
vector $\boldsymbol{\Gamma}$ is the magnetoelectric coupling
parameter coupling the electric and magnetic fields. Metamaterials
are artificial materials made of sub-wavelength metallic
constituents which are randomly or periodically distributed in a
dielectric background. In some metamaterials the cross
polarization effect (an electric polarization results from an
applied magnetic field and vice versa) occurs which are called
bianisotropic or isotropic metamaterials \cite{bianisotropic1}. By
choosing suitable distribution of the metallic inclusions in these
materials, it is possible to obtain desired electric permittivity,
magnetic permeability and magnetoelectric coupling parameter
following Eq. (\ref{eq:epsilonmu}). Fabrication of bianisotropic
and biisotropic metamaterials operating in the visible wavelengths
is possible using different techniques such as lithography and
laser writing \cite{bianisotropic1}. To understand how light wave
propagates near a black hole in different space-times, we can
experimentally and theoretically study the propagation of
electromagnetic waves in the corresponding metamaterials. Then, to
prove the correctness of this equivalence, we compare our results
with trajectories obtained using Lagrangian formalism. In the
other words, since the trajectories near black holes in different
spacetimes are well studied, we compare them with the trajectories
obtained from the equivalence optical medium to prove the
correctness of our method. It should be noted that the
electromagnetic (optical) black hole \cite{Opt-BHs} is fabricated
for the first time in 2010 based on metamaterials composed of
nonresonant I-shaped inclusions and electric field coupled
resonators \cite{obh2}. This structure can absorb the
electromagnetic wave incident from every direction like a black
hole or a black body.

In other words, the mixing of electric and magnetic fields is
addressed by $\boldsymbol{\Gamma}$ with the physical dimension of
velocity. It is also shown that for a slow-moving medium, $u/c
<<1$, $\boldsymbol{\Gamma}$ is proportional to the velocity of
medium \cite{velocity}.

%%%%%%%%%%%%%%%%%%%%%%%%%%%%%%%%%%%%%%%%%%%%%%%%%%%%%%%%%%%%%%%%%%%%%%%%%%%%%%%%%%%%%%%%%%%%%

\section{Maxwell equations in medium}
Regarding to the first approach, Maxwell equations written in a
source-free form \cite{jackson}
\begin{eqnarray}
\nabla . \vec{D}=0,\hspace{.5cm}
\nabla \times \vec{H}=\frac{\partial \vec{D}}{\partial t}, \hspace{.5cm}
\nabla \times \vec{E}=-\frac{\partial \vec{B}}{\partial t}, \hspace{.5cm}
\nabla .\vec{B}=0,
\label{eq:maxwell}
\end{eqnarray}
should be supplemented by the constitutive relations Eq.
(\ref{eq:plebanski}) and medium tensors expressed spacetime, in
order to be solved. Since the constitutive equations couple the
magnetic and electric fields, Eq. (\ref{eq:maxwell}) become too
complicated to be solved, we have to seek for a method to decouple
equations. In this regard, we apply the time-harmonic Maxwell
equations, assuming a time dependence $e^{-i\omega t}$ with $t$
and $\omega$ as time variable and angular wave frequency,
respectively,
\begin{eqnarray}
\nabla . \vec{D}=0,\hspace{.5cm}
\nabla \times \vec{H}=-i\omega \vec{D}, \hspace{.5cm}
\nabla \times \vec{E}=i\omega \vec{B}, \hspace{.5cm}
\nabla .\vec{B}=0,
\label{eq:thmaxwell}
\end{eqnarray}
where $\vec{E}$, $\vec{H}$, $\vec{D}$ and $\vec{B}$ are,
respectively, electric field, magnetic field, electric
displacement and magnetic flux density. We have considered that
the constitutive equations obey Eq. (\ref{eq:plebanski}), and
therefore, substituting this equation to Eq. (\ref{eq:thmaxwell})
we can find the following Maxwell-Tellegen equations \cite{fea}:
\begin{eqnarray}
\nabla \times \vec{H}=-i\omega \epsilon \vec{E} + i\omega \vec{\Gamma}\times \vec{H}, \hspace{.5cm}
\nabla \times \vec{E}=i\omega \mu \vec{H}+ i\omega \vec{\Gamma}\times \vec{E}. \hspace{.5cm}
\label{eq:fea}
\end{eqnarray}

\par
We have assumed that the material parameters, $\epsilon$, $\mu$
and $\Gamma$, and the electromagnetic fields are invariant in one
direction, here $z$ direction is considered. Since $\epsilon$,
$\mu$ and $\Gamma$ are $z-$anisotropic tensors, in general, we can
write:
\begin{eqnarray}
\epsilon^{ij}=
\begin{bmatrix}
\epsilon_{11} & \epsilon_{12} & 0 \\
\epsilon_{21} & \epsilon_{22} & 0 \\
0 & 0 & \epsilon_{33}
\end{bmatrix}, \hspace{0.5cm}
\mu^{ij}=
\begin{bmatrix}
\mu_{11} & \mu_{12} & 0 \\
\mu_{21} & \mu_{22} & 0 \\
0 & 0 & \mu_{33}
\end{bmatrix},
\hspace{0.5cm}
\vec{\Gamma}=
\begin{bmatrix}
\Gamma_{1} \\
\Gamma_{2}  \\
0
\end{bmatrix}
\label{eq:et}.
\end{eqnarray}

Applying Eq. (\ref{eq:et}) to Eq. (\ref{eq:fea}) leads to:
\begin{eqnarray}
-\partial_x E_3=i\omega (\mu_{21}H_1+\mu_{22}H_2)-i\omega
\Gamma_1E_3, \label{eq:elecmegfield1}
\end{eqnarray}
\begin{eqnarray}
\partial_y E_3=i\omega (\mu_{11}H_1+\mu_{12}H_2)+i\omega \Gamma_2E_3, \label{eq:elecmegfield2}
\end{eqnarray}
\begin{eqnarray}
\partial_x E_2-\partial_y E_1=i\omega \mu_{33}H_3+i\omega(\Gamma_1E_2-\Gamma_2 E_1), \label{eq:elecmegfield3}
\end{eqnarray}
\begin{eqnarray}
\partial_x H_3=i\omega (\epsilon_{21}E_1+\epsilon_{22}E_2)+i\omega \Gamma_1H_3, \label{eq:elecmegfield4}
\end{eqnarray}
\begin{eqnarray}
\partial_y H_3=-i\omega (\epsilon_{11}E_1+\epsilon_{12}E_2)+i\omega \Gamma_2H_3, \label{eq:elecmegfield5}
\end{eqnarray}
\begin{eqnarray}
\partial_x H_2-\partial_y H_1=-i\omega \epsilon_{33}E_3+i\omega(\Gamma_1H_2-\Gamma_2 H_1). \label{eq:elecmegfield6}
\end{eqnarray}

It is patently obvious from Eq.
(\ref{eq:elecmegfield1}-\ref{eq:elecmegfield6}) that electric and
magnetic fields are coupled because of the constitutive equations.
In order to decouple these equations, we follow the approach
introduced in Ref. \cite{fea}. At the first step, we rewrite Eqs.
(\ref{eq:elecmegfield1}) and (\ref{eq:elecmegfield2}) as the
following unified relation
\begin{eqnarray}
\underline{A} E_3=i\omega \mu_T \underline{H}+i\omega \Gamma_T E_3, \label{eq:em1}
\end{eqnarray}
and also Eqs. (\ref{eq:elecmegfield4}) and
(\ref{eq:elecmegfield5}) as
\begin{eqnarray}
\underline{A} H_3=i\omega \epsilon_T \underline{E}+i\omega \Gamma_T H_3\label{eq:em2},
\end{eqnarray}
where
\begin{equation}
\begin{array}{l}
\underline{A}=\begin{bmatrix}
\partial_x \\
\partial_y
\end{bmatrix},\hspace{0.5cm} \underline{E}=\begin{bmatrix}
E_2 \\
-E_1
\end{bmatrix},\hspace{0.5cm} \underline{H}=\begin{bmatrix}
H_2 \\
-H_1
\end{bmatrix}, \hspace{0.5cm} \Gamma_T=\begin{bmatrix}
\Gamma_1 \\
\Gamma_2
\end{bmatrix},
\\
\\
\\
\epsilon_T=\begin{bmatrix}
\epsilon_{22} & -\epsilon_{12} \\
-\epsilon_{21} & \epsilon_{11}
\end{bmatrix}, \hspace{0.5cm} \mu_T=\begin{bmatrix}
-\mu_{22} & \mu_{12} \\
\mu_{21} & -\mu_{11}
\end{bmatrix}.
\end{array}\label{eq:pdeco}
\end{equation}

Next, one can drive $\underline{H}$ and $\underline{E}$ from Eqs.
(\ref{eq:em1}) and (\ref{eq:em2}), respectively
\begin{eqnarray}
\underline{H}=(i\omega \mu_T)^{-1}(\underline{A}-i\omega \Gamma_T)E_3, \label{eq:h}
\end{eqnarray}
and
\begin{eqnarray}
\underline{E}=(i\omega \epsilon_T)^{-1}(\underline{A}-i\omega \Gamma_T)H_3. \label{eq:e}
\end{eqnarray}
We can also rewrite Eqs. (\ref{eq:elecmegfield3}) and
(\ref{eq:elecmegfield6}) in the following forms
\begin{eqnarray}
\begin{bmatrix}
\partial_x & \partial_y
\end{bmatrix}\underline{E}=i\omega \mu_{33}H_3+i\omega
\begin{bmatrix}
\Gamma_1 & \Gamma_2
\end{bmatrix}\underline{E},
\label{eq:e2}
\end{eqnarray}
and
\begin{eqnarray}
\begin{bmatrix}
\partial_x & \partial_y
\end{bmatrix}\underline{H}=-i\omega \epsilon_{33}E_3+i\omega
\begin{bmatrix}
\Gamma_1 & \Gamma_2
\end{bmatrix}\underline{H}.
\label{eq:h2}
\end{eqnarray}

Substituting Eqs. (\ref{eq:e}) and (\ref{eq:h}) into Eqs.
(\ref{eq:e2}) and (\ref{eq:h2}), one finds
\begin{equation}
\begin{array}{l}
\begin{bmatrix}
\partial_x & \partial_y
\end{bmatrix}(i\omega \epsilon_T)^{-1}(\underline{A}-i\omega\Gamma_T)H_3=
\\
\hspace{4cm}i\omega\mu_{33}H_3
+i\omega\begin{bmatrix}
\partial_x & \partial_y
\end{bmatrix}(i\omega\epsilon_T)^{-1}(\underline{A}-i\omega\Gamma_T)H_3
\end{array}\label{eq:h3},
\end{equation}
and
\begin{equation}
\begin{array}{l}
\begin{bmatrix}
\partial_x & \partial_y
\end{bmatrix}(i\omega \mu_T)^{-1}(\underline{A}-i\omega\Gamma_T)E_3=
\\
\hspace{4cm}-i\omega\epsilon_{33}E_3
+i\omega\begin{bmatrix}
\partial_x & \partial_y
\end{bmatrix}(i\omega\mu_T)^{-1}(\underline{A}-i\omega\Gamma_T)E_3
\end{array}\label{eq:e3}.
\end{equation}

It is notable that all Eqs.
(\ref{eq:elecmegfield1}-\ref{eq:elecmegfield6}) are resolved to
two Eqs. (\ref{eq:h3}) and (\ref{eq:e3}), which are decoupled
equations with the longitudinal fields $E_3$ and $H_3$ as
unknowns.

We have considered TE polarization for an electromagnetic wave for
which $\vec{E}$ is perpendicular to the $xy$ plane. If $\vec{E}$
is determined in the $z$ direction, the only non-zero component is
$E_3$. Hence, as the electric and magnetic fields are
perpendicular, $H_3$ is zero. By applying these assumptions to
Eqs. (\ref{eq:h3}) and (\ref{eq:e3}), one concludes that only Eq.
(\ref{eq:e3}) is required to solve, since Eq. (\ref{eq:h3}) is
always satisfied.

\section{Lagrangian formalism}

In the previous section, we used wave optics in order to determine
the propagation of light in  optical materials mimicking
spacetime. In this section, we apply geometrical optics in order
to determine the propagation of ray in that materials to be assure
that our model works well. In this regard, the Lagrangian is
introduced, then based on the equation of motion the ray path is
obtained.

It was shown \cite{mtobh} that the equations governing the
geodesics in  a spacetime with the line element $ds^2=g_{\alpha
\beta}dx^{\alpha}dx^{\beta}$ can be derived from the Lagrangian
\begin{eqnarray}
\mathcal{L}=g_{\alpha \beta}\frac{dx^{\alpha}}{d\tau}
\frac{dx^{\beta}}{d\tau}=g_{\alpha
\beta}\dot{q}^{\alpha}\dot{q}^{\beta}.\label{eq:lagrangian}
\end{eqnarray}
where $\tau$ is the affine parameter along the geodesics. From the
calculus of variations, one can write the following
Euler--Lagrange equation
\begin{eqnarray}
\frac{\partial \mathcal{L}}{\partial q^{\alpha}}-\frac{d}{d\tau}
\frac{\partial \mathcal{L}}{\partial
\dot{q^{\alpha}}}=0\label{eq:EulerLagrange},
\end{eqnarray}
where $q^{\alpha}=(t,r,\theta,\phi)$ is the generalized
coordinate, and the generalized momentum $p_{\alpha}$ is defined
as
\begin{eqnarray}
p_{\alpha}=\frac{\partial \mathcal{L}}{\partial \dot{q^{\alpha}}}.
\label{eq:pi}
\end{eqnarray}
It is easy to determine $\dot{p_i}$ from the Euler--Lagrange
equation, Eq. (\ref{eq:EulerLagrange}), as below
 \begin{eqnarray}
\dot{p_{\alpha}}=\frac{\partial \mathcal{L}} {\partial
q^{\alpha}}\label{eq:pdoti},
\end{eqnarray}
where dot denotes the derivative over the affine parameter,
$\tau$. Based on Eqs. (\ref{eq:lagrangian}) and (\ref{eq:pi}), we
can rewrite the Lagrangian as a function of generalized momentum,
$p_i$. Since we are seeking the ray path, the Lagrangian, Eq.
(\ref{eq:lagrangian}), is zero.

\section{Reissner-Nordstr\"{o}m Spacetime}

First, we apply the analogy to the Reissner-Nordstr\"{o}m
spacetime. The Reissner-Nordstr\"{o}m metric is a static solution
of the Einstein-Maxwell field equations, which corresponds to the
gravitational field of a charged, non-rotating, spherically
symmetric body of mass $M$ \cite{inverno}. This metric in the
Cartesian coordinates can be written as
\begin{equation}
\begin{array}{l}
ds^2=(\frac{2m}{r}-1-\frac{Q^2}{r^2})dt^2+
(\frac{x^2}{r^2-2mr+Q^2}+\frac{x^2
z^2}{r^2(r^2-z^2)}+\frac{y^2}{r^2-z^2})dx^2
\\
+(\frac{y^2}{r^2-2mr+Q^2}+\frac{y^2 z^2}{r^2(r^2-z^2)}
+\frac{x^2}{r^2-z^2})dy^2+(\frac{z^2}{r^2-2mr+Q^2}+\frac{r^2-z^2}{r^2})dz^2
\\
+2(\frac{xy}{r^2-2mr+Q^2}-\frac{xy}{r^2})dxdy+
2(\frac{zx}{r^2-2mr+Q^2}-\frac{zx}{r^2})dxdz
\\
+2(\frac{zy}{r^2-2mr+Q^2}-\frac{zy}{r^2})dydz.
\end{array} \label{eq:rncartesian}
\end{equation}

Here $Q$ is the charge of black hole, $m$ is the geometrical mass
of the source of gravitation and $r=\sqrt{x^2+y^2+z^2}$.

The anisotropic equivalent medium tensors can be obtained from Eq.
(\ref{eq:epsilonmu}). It is notable that $\Gamma_i=0$ because the
Reissner-Nordstr\"{o}m metric is static. Therefore, permittivity
and permeability tensors in the $xy$ plane can be obtained as
\begin{eqnarray}
\epsilon^{ij}=\mu^{ij}=
    \setlength{\tabcolsep}{3pt}
    \renewcommand{\arraystretch}{3}
    \begin{pmatrix}
    \frac{r^4+Q^2x^2-2mrx^2}{r^2(r^2+Q^2-2mr)} & \frac{xy(Q^2-2mr)}{r^2(r^2+Q^2-2mr)} & 0\\
    \frac{xy(Q^2-2mr)}{r^2(r^2+Q^2-2mr)} & \frac{r^4+Q^2y^2-2mry^2}{r^2(r^2+Q^2-2mr)} & 0   \\
    0 & 0 & \frac{r^2}{(r^2+Q^2-2mr)}.
    \end{pmatrix}.
\label{eq:rnconsz}
\end{eqnarray}

Since we are working in the $xy$ plane, here $r=\sqrt{x^2+y^2}$.
Having permittivity and permeability tensors, Eq. (\ref{eq:e3})
can be solved. We have solved the wave equation Eq. (\ref{eq:e3})
numerically for a beam with the appropriate boundary conditions.

It is known that having a non-trivial spherically symmetric
electromagnetic wave is not possible. Its reason comes from the
fact that the polarization vectors of such a field configuration
would introduce a continuous nowhere-vanishing vector field
tangent to the 2-sphere, thereby contradicting the well-known fact
that the 2-sphere is not parallelizable, and therefore, we
motivate to use other coordinate systems. The optical parameters
of the material equivalent to the black hole depend on the
coordinate system. Because working with Cartesian coordinate in
COMSOL is easier than other coordinate systems, we use it in this
paper. It is also notable that although the (non-physical)
coordinate singularity can be removed in the Cartesian coordinate,
its one-way membrane property is kept for the equivalent
metamaterials.

The Reissner-Nordstr\"{o}m metric in spherical coordinates reads
\cite{inverno}
\begin{equation}
ds^2=-(1-\frac{2m}{r}+\frac{Q^2}{r^2})dt^2+
(1-\frac{2m}{r}+\frac{Q^2}{r^2})^{-1}dr^2+
r^2(d\theta^2+sin^2\theta d\phi^2). \label{eq:rn}
\end{equation}

Substituting this metric into Eq. (\ref{eq:lagrangian}) and
considering the $xy$ plane, we find
\begin{equation}
\mathcal{L}=-(1-\frac{2m}{r}+\frac{Q^2}
{r^2})\dot{t}^2+\frac{\dot{r}^2}
{(1-\frac{2m}{r})+\frac{Q^2}{r^2}}+ {r^2}\dot{\phi}^2.
\label{eq:hamiltonrn}
\end{equation}

Considering the obtained Lagrangian and using Eqs. (\ref{eq:pi})
and (\ref{eq:pdoti}), $p_i$ and $\dot{p_i}$ can be easily
calculated as
\begin{eqnarray}
\dot{p_{t}}=\frac{dp_{t}}{d\tau}=0\label{eq:pdottrn},
\end{eqnarray}
\begin{eqnarray}
\dot{p_{\phi}}=\frac{dp_{\phi}}{d\tau}=0\label{eq:pdotphirn},
\end{eqnarray}
\begin{eqnarray}
p_t=-2(1-\frac{2m}{r}+\frac{Q^2}{r^2})\dot{t}\label{eq:ptrn},
\end{eqnarray}
\begin{eqnarray}
p_{\phi}=2r^2\dot{\phi}. \label{eq:phidotrn}
\end{eqnarray}

We define $p_t=2E$ and $p_{\phi}=2L$ that are constant since
$\dot{p_t}=0$ and $\dot{p_{\phi}}=0$, which just state the
conservation of energy and angular momentum for the static
spherically symmetric system. Hence, the Lagrangian Eq.
(\ref{eq:hamiltonrn}) can be rewritten as
\begin{eqnarray}
\mathcal{L}=\frac{-E^2}{(1-\frac{2m}{r} +\frac{Q^2}{r^2})}
+\frac{\dot{r}^2}{(1-\frac{2m}{r}+
\frac{Q^2}{r^2})}+\frac{L^2}{r^2}.  \label{eq:hamiltoneqrn}
\end{eqnarray}

Considering $\mathcal{L}=0$, one finds
\begin{eqnarray}
\dot{r}^2=(\frac{dr}{d\tau})^2=E^2-\frac{L^2}{r^2}
(1-\frac{2m}{r}+\frac{Q^2}{r^2})=f(r), \label{eq:drdtaurn}
\end{eqnarray}
and $\dot{\phi}$ can be determined from Eq. (\ref{eq:phidotrn}),
therefore
\begin{eqnarray}
(\frac{dr}{d\phi})^2=\frac{r^4}{D^2}- r^2+2mr-Q^2.
\label{eq:drdphirn}
\end{eqnarray}

Since photons are massless, the ray trajectories can be determined
by only one character which is called the impact parameter
$D=\frac{L}{E}$ \cite{mtobh}. The ray trajectories can be obtained
by solving Eq. (\ref{eq:drdphirn}) \cite{hakmann,mtobh}. The
results for Reissner-Nordstr\"{o}m spacetime, both for wave optics
and ray optics are shown in Figs. 1-4.

Interestingly, depending on the roots of $f(r)$, we have three
types of trajectories: "terminating orbit", "bound orbit" and
"flyby orbit". Those roots depend only on the impact parameter,
$D$, since both $m$ and $Q$ are constants. Therefore, for $D=D_c$
we have "bound orbit", if $D>D_c$ there is "flyby orbit" and if
$D<D_c$ there is "terminating orbit" \cite{hakmann}. The critical
value of impact parameter, $D_c$, can be driven by considering
$f^{\prime}(r_c)=0$ and $f(r_c)=0$
\begin{equation}
r_c=\frac{3}{2}m\big[1+\sqrt{1-\frac{8Q^2}{9m^2}}\big],
\hspace{0.5cm} D_c=\frac{r_c^2}{\sqrt{r_c^2-2mr_c+Q^2}}.
\label{eq:dcrcrn}
\end{equation}

It is obvious that for different values of charge, $Q$, the
critical value of impact parameter differs.

\section{Kerr Spacetime}

In this section, we are going to generalize the Schwarzschild
static solution to the case of stationary one, and therefore, the
Kerr spacetime is discussed. The Kerr metric describes the
geometry of empty spacetime around a rotating uncharged
axially-symmetric black hole with an appropriate event horizon.
The Kerr metric is an exact solution of the Einstein field
equations of general relativity with stationary constraint
\cite{inverno}. This metric in the Cartesian coordinates reads
\begin{eqnarray}
ds^2=d\bar{t}^2-dx^2-dy^2-dz^2-
\frac{2mr^3}{r^4+a^2z^2}(d\bar{t}+\frac{r(xdx+ydy)}
{a^2+r^2}+\frac{a(ydx-xdy)}{a^2+r^2}+\frac{zdz}{r})^2,
\label{eq:kerrcartesian}
\end{eqnarray}
where $a$ and $m$ are spin parameter and geometrical mass,
respectively, the radial coordinate is $r=\sqrt{x^2+y^2+z^2}$ and
also $d\bar{t}=dt+\frac{2m(xdx+ydy+zdz)}{r^2-2mr+a^2}$.

Based on Eq. (\ref{eq:epsilonmu}), permittivity and permeability
tensors and $\Gamma_i$ in the $xy$ plane can be obtained as
 \begin{equation}
 \begin{array}{l}
 \epsilon^{xx}=\mu^{xx}=-A_z[\frac{-r^8+2mr^7-2r^6a^2+
 4m(a^2-\frac{y^2}{2})r^5+(-a^4+4mxya)r^4+2ma^2(a^2-x^2)r^3}
 {(1-\frac{2m}{r})(a^2+r^2)(r^6-2mr^5+r^4a^2-2m(a^2-r^2)r^3)}],
 \\
 \\
 \epsilon^{xy}=\epsilon^{yx}=\mu^{xy}=\mu^{yx}=2A_z[\frac{(ay+rx)
 (ax-ry)mr^3}{(1-\frac{2m}{r})(a^2+r^2)(r^6-2mr^5+r^4a^2-2m(a^2-r^2)r^3)}],
 \\
 \\
 \epsilon^{xz}=\epsilon^{zx}=\mu^{xz}=\mu^{zx}=0,
 \\
 \\
 \epsilon^{yy}=\mu^{yy}=-A_z[\frac{-r^8+2mr^7-2r^6a^2+4m(a^2-
 \frac{x^2}{2})r^5+(-a^4-4mxya)r^4+2ma^2(a^2-y^2)r^3}{(1-\frac{2mr}
 {r})(a^2+r^2)(r^6-2mr^5+r^4a^2-2m(a^2-r^2)r^3)}],
 \\
 \\
 \epsilon^{zy}=\epsilon^{yz}=\mu^{zy}=\mu^{yz}=0,
 \\
 \\
 \epsilon^{zz}=\mu^{zz}=-A_z[\frac{-r^6+2mr^5-r^4a^2+2m(a^2-x^2-y^2)r^3}
 {(1-\frac{2m}{r})(a^2+r^2)(r^6-2mr^5+r^4a^2-2m(a^2-r^2)r^3)}],
 \end{array}\label{eq:kerrconsz0}
 \end{equation}
where $A_z=\sqrt{\frac{r^4a^2-2mr^3a^2+r^6}{r^4(a^2+r^2)}}$. In
addition, we find that
 \begin{equation}
 \begin{array}{l}
 \Gamma_x=-\frac{2ma(-2mr^3xa-r^5y+2r^4ym-r^3a^2y)}{r^4(r^2-2mr+a^2)(a^2+r^2)(1-\frac{2mr}{r})},
 \\
 \\
 \Gamma_y=-\frac{2ma(-2mr^3ya+r^5x-2r^4xm+r^3a^2x)}{r^4(r^2-2mr+a^2)(a^2+r^2)(1-\frac{2m}{r})},
 \\
 \\
 \Gamma_z=0.
 \end{array}\label{eq:kerrgammaz0}
 \end{equation}

It is interesting that the vector $\vec{\Gamma}$ is non-zero
because the Kerr metric is not static. It can be said that the
spin parameter, $a$, makes the equivalent material bi-anisotropic.
In addition, Eq. (\ref{eq:e3}) can be solved numerically by
considering Eqs. (\ref{eq:kerrconsz0}) and (\ref{eq:kerrgammaz0}).

Regarding the Lagrangian formalism, first, we write the Kerr
metric in the spherical coordinates \cite{inverno}
\begin{equation}
ds^2=\frac{\Delta}{\rho^2}(dt-asin^2\theta d\phi)^2-\frac{sin^2\theta}{\rho^2}[(r^2+a^2)d\phi-adt]^2
-\frac{\rho^2}{\Delta}dr^2-\rho^2d\theta^2,
\label{eq:kerr}
\end{equation}
where $\rho^2=r^2+a^2cos^2\theta$ and $\Delta=r^2-2mr+a^2$. Since
we are working in the $xy$ plane, we set $\theta=\frac{\pi}{2}$.
Substituting this metric into Eq. (\ref{eq:lagrangian}), one can
show that
\begin{equation}
\mathcal{L}=(1-\frac{2ma^2}{r})\dot{t}^2+
(\frac{4ma}{r})\dot{t}\dot{\phi}-\frac{r^2}{r^2-2mr+a^2}\dot{r}^2
-[r^2+a^2+\frac{2ma^2}{r}]\dot{\phi}^2. \label{eq:kerrlagrangian}
\end{equation}

Using Eqs. (\ref{eq:pi}) and (\ref{eq:pdoti}), the following
relations are obtained
\begin{eqnarray}
\dot{p_{t}}=\frac{dp_{t}}{d\tau}=0,   \label{eq:pdottkerr}
\end{eqnarray}
\begin{eqnarray}
\dot{p_{\phi}}=\frac{dp_{\phi}}{d\tau}=0  \label{eq:pdotphikerr}
\end{eqnarray}
\begin{eqnarray}
p_t=2(1-\frac{2m}{r})\dot{t}+\frac{4ma^2}{r}\dot{\phi},
\label{eq:ptkerr}
\end{eqnarray}
\begin{eqnarray}
p_{\phi}=\frac{4ma}{r}\dot{t}-2(r^2+a^2+\frac{2ma^2}{r})\dot{\phi}.
\label{eq:pphikerr}
\end{eqnarray}
where, we call $p_t=2E$ and $p_{\phi}=-2L$, as before. By solving
Eqs. (\ref{eq:ptkerr}) and (\ref{eq:pphikerr}) simultaneously,
$\dot{t}$ and $\dot{\phi}$ can be calculated as
\begin{eqnarray}
\dot{t}=\frac{1}{r^2-2mr+a^2}[(1-\frac{2m}{r})L+\frac{2ma}{r}E],
\label{eq:tdotkerr}
\end{eqnarray}
\begin{eqnarray}
\dot{\phi}=\frac{1}{r^2-2mr+a^2}[(r^2+a^2+\frac{2ma^2}{r})
E-\frac{2ma}{r}L]. \label{eq:phidotkerr3}
\end{eqnarray}

Moreover, substituting these parameters into the Lagrangian Eq.
(\ref{eq:kerrlagrangian}), one finds
\begin{eqnarray}
\mathcal{L}=(a^2+r^2+\frac{2ma^2}{r})E^2+(\frac{2m}{r})
L^2-\frac{4ma}{r}LE-r^2\dot{r}^2. \label{eq:kerrhamiltonian}
\end{eqnarray}

Since for light beam we have $\mathcal{L}=0$,
\begin{eqnarray}
\dot{r}^2=E^2+\frac{2m}{r^3}(aE-L)^2+\frac{1}{r^2}(a^2E^2-L^2)=f(r).
\label{eq:rdotkerr}
\end{eqnarray}

It is noteworthy to mention that only the impact parameter affects
the ray trajectories which its effects lead to three types of
orbits mentioned in the previous section. At this point, since the
equations related to Kerr spacetime are too complicated, we just
focus on critical value of impact parameter. In this regard, by
calculating $f^{\prime}(r_c)=0$ and $f(r_c)=0$, one finds
\cite{mtobh}
\begin{equation}
r_c=3m\frac{L-aE}{L+aE}=3m\frac{D_c-a}{D_c+a}, \hspace{0.5cm}
(D_c+a)^3=27m^2(D_c-a). \label{eq:dcrckerr}
\end{equation}

It is clear that the spin parameter, $a$, controls the critical
value of the impact parameter. By considering $u=\frac{1}{r}$ and
considering relations in  Eqs. (\ref{eq:rdotkerr}) and
(\ref{eq:dcrckerr}), we can rewrite Eq. (\ref{eq:rdotkerr}) as
\begin{eqnarray}
\dot{u}^2=E^2u^4m(D_c-a)^2(u-u_c)^2(2u+u_c). \label{eq:udotkerr}
\end{eqnarray}

Taking into account $\dot{\phi}$ in Eq. (\ref{eq:phidotkerr3}), we
can obtain
\begin{eqnarray}
\frac{du}{d\phi}=\frac{(1-2mu+a^2u^2)(u-u_c)(D_c+a)\sqrt{2u+u_c}}{\sqrt{m}[3u_cD_c-2u(D_c+a)]}.
\label{eq:dudphikerr}
\end{eqnarray}

Furthermore, solving Eq. (\ref{eq:dudphikerr}) gives the ray
trajectories \cite{mtobh}. The results for Kerr spacetime are
shown in Fig. 5.

\section{Results and discussion}

Here, we are in a position to solve the equations of two
approaches and simulate the results for comparison. Equation
(\ref{eq:e3}) is solved in the Cartesian coordinates and a
rectangular $2D$ geometry of $xy$ space for a $TE$ polarized wave
of frequency $\omega=6.3\times 10^9(rad/s)$ injected from the
right and the results are compared with the ray path (red line)
calculated from the equations of motion. The computational domain
is surrounded by a perfectly matched layer that absorbs the
outward waves to ensure that there are no unwanted reflections,
and the simulations are done by means of a standard software
solver (COMSOL).

For Reissner-Nordstr\"{o}m spacetime the medium parameters are
given by Eq. (\ref{eq:rnconsz}) and the results are depicted in
Table \ref{tabrn} and Figs. 1-4. In order to obtain dimensionless
parameters, we normalize all parameters to $r_0=\frac{r_s}{2}$ in
which $r_s=2m$ is the event horizon of Schwarzschild black hole
where the gravity is so strong, that light cannot escape
\cite{mtobh}.

%%%%%%%%%%%%%%%%%%%%%%%%%%%%%%%%%%%%%%%%%%
\begin{table}[H]
    \begin{center}
        \begin{tabular}{|c|c|c|}
            \hline\hline $Q$ & $D_{c}$ & $r_c$\\
            \hline \hline $0$ & $5.196$ & $3$\\
            \hline $0.1$ & $5.187$ & $2.99$\\
            \hline
            $0.5$ & $4.97$ & $2.82$\\
            \hline$1$ & $4$ & $2$\\ \hline
        \end{tabular}%
        \\[0pt]
        \caption{Reissner-Nordstr\"{o}m solutions: the numerical results
        of $D_c$ and $r_c$ for different values of $Q$ in the case of $m=1$. } \label{tabrn}
    \end{center}
\end{table}
%%%%%%%%%%%%%%%%%%%%%%%%%%%%%%%%%%%%%%%%%%

\begin{figure}[H]
    \centering
    \subfloat[$D=3.2$]{\includegraphics[width=15cm]{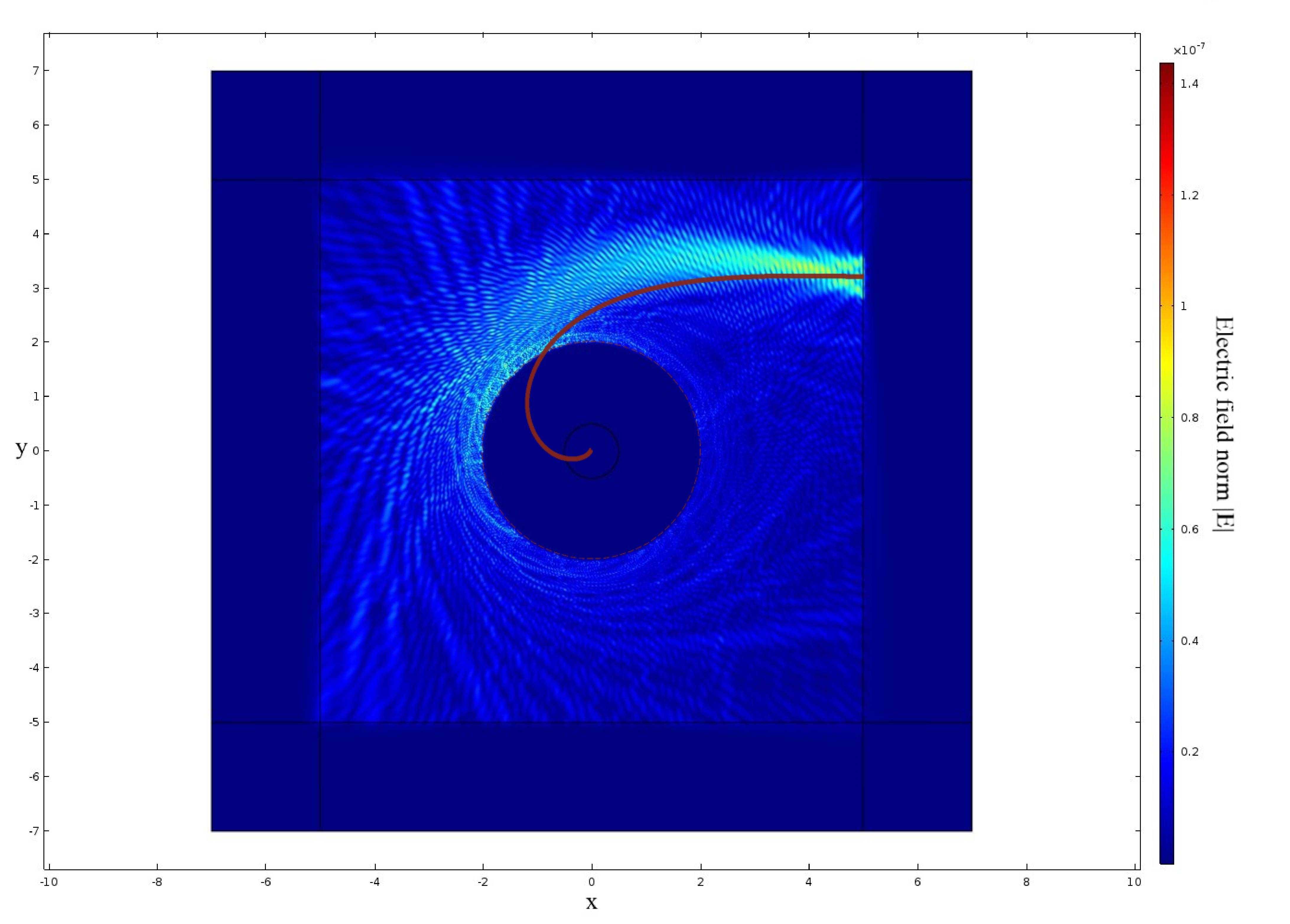}\label{schwarzm1D32final}}
    \hfill
    \subfloat[$D=5.2$]{\includegraphics[width=15cm]{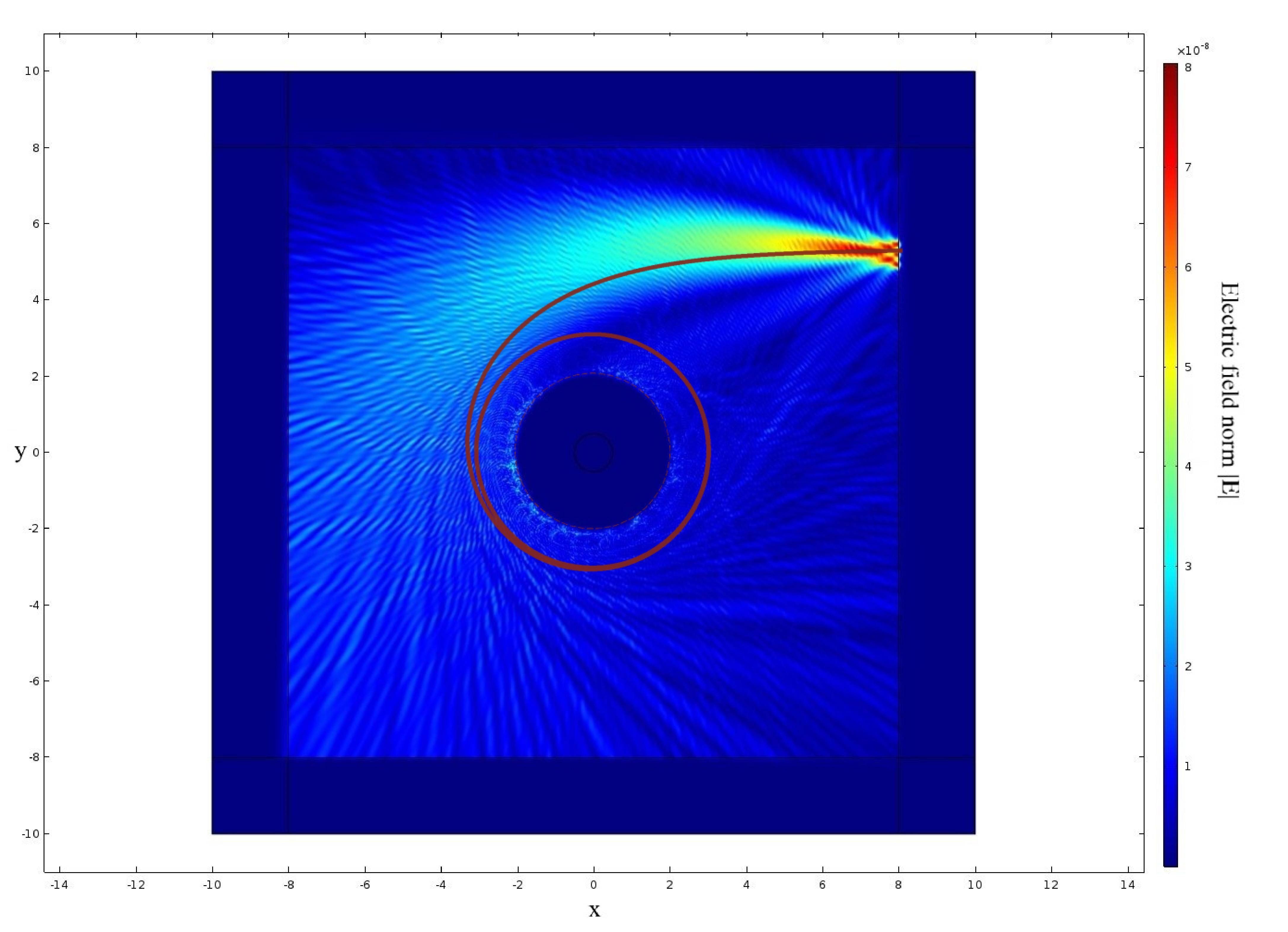}
        \label{schwarzm1D52final}}
\end{figure}

\begin{figure}
        \centering
    \subfloat[][$D=6.2$]{\includegraphics[width=15cm]{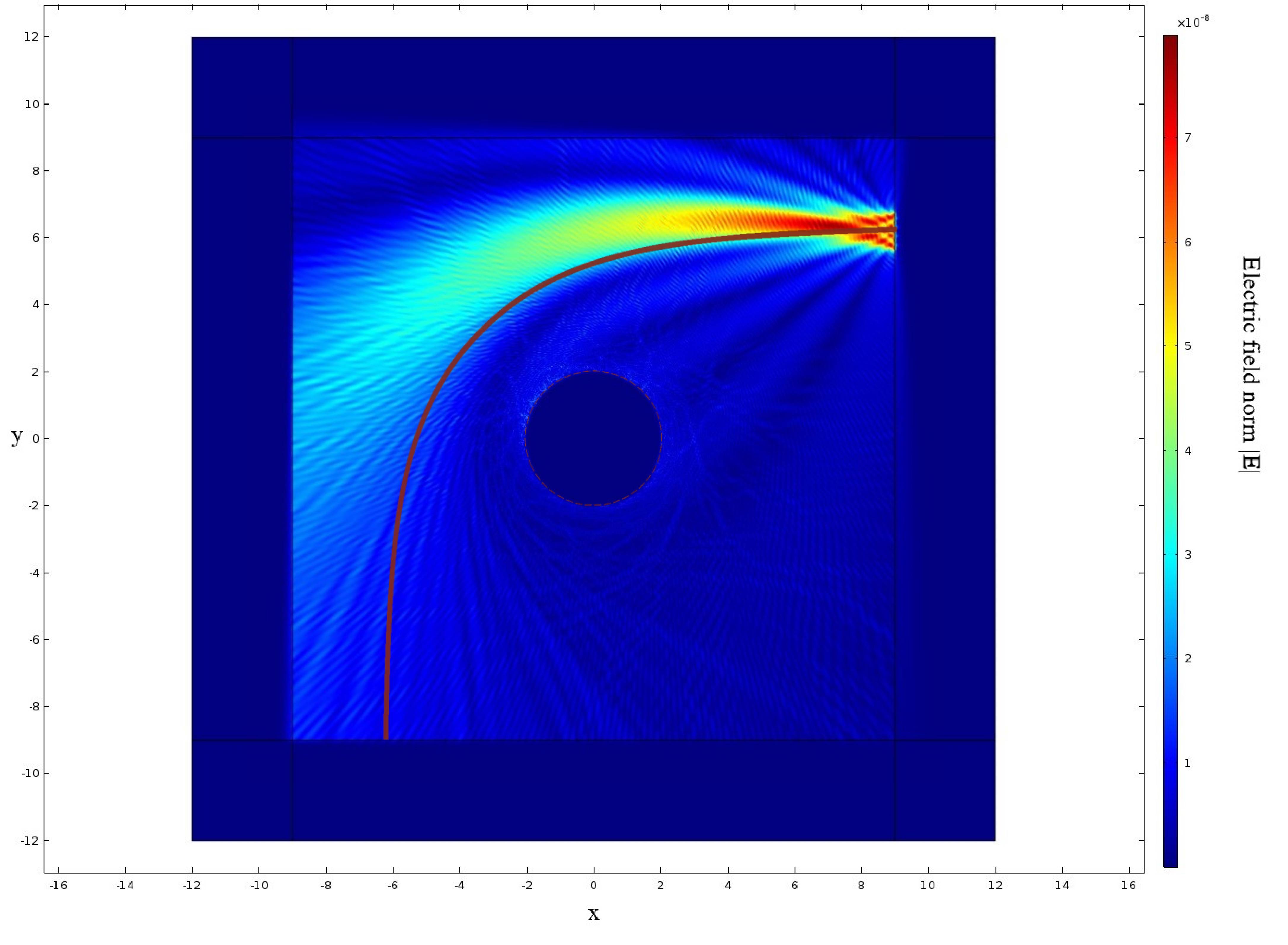}\label{schwarzm1D62final}}
    \label{schwarz} \caption{metamaterial mimicking the Reissner-Nordstr\"{o}m spacetime
        for $Q=0$, $m=1$ and different impact parameters, $D$.}
\end{figure}

\begin{figure}[H]
    \centering
    \subfloat[$D=3.2$]{\includegraphics[width=15cm]{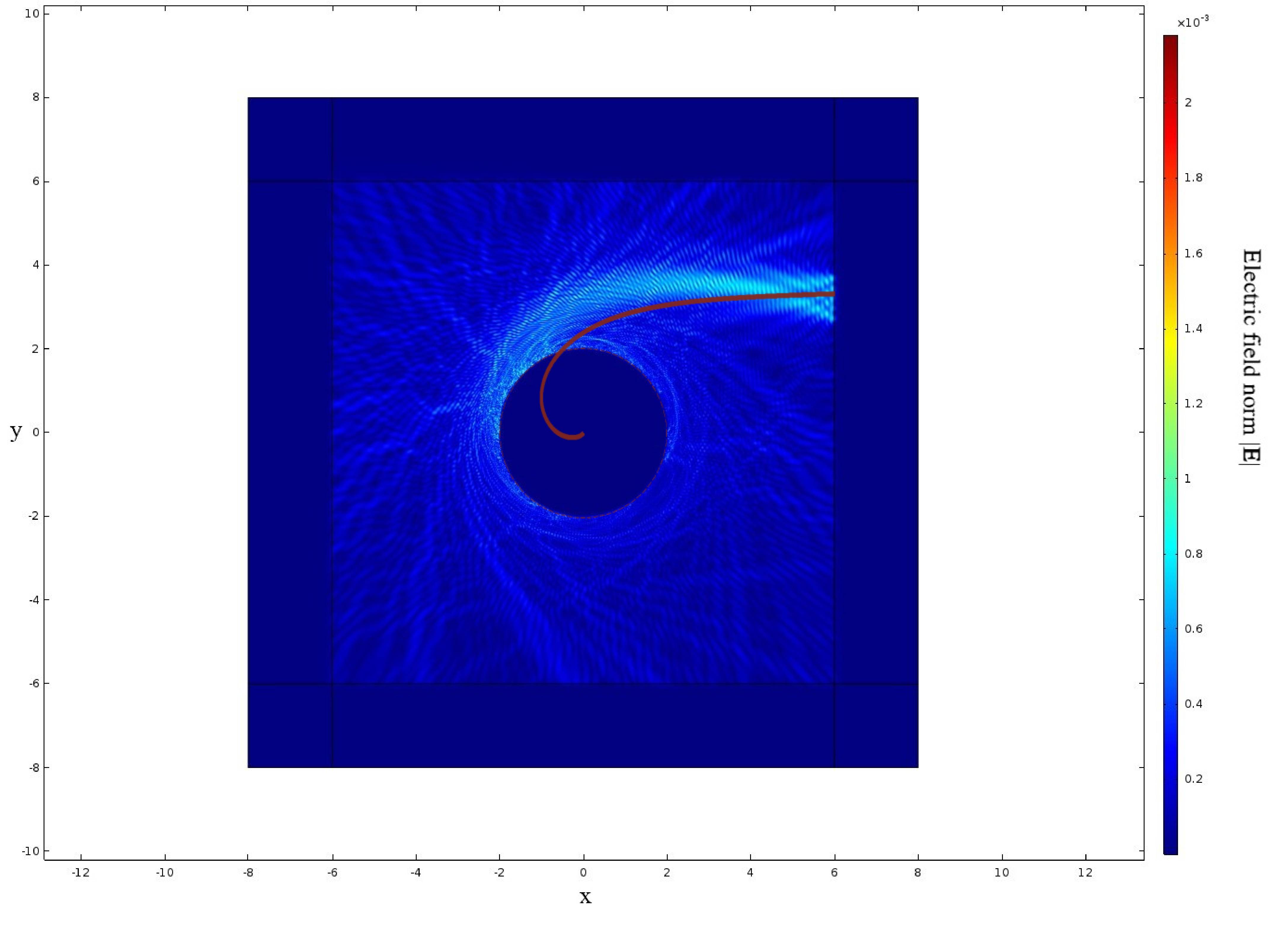}
        \label{RNm1Q1D32final}}
    \hfill
    \subfloat[$D=5.2$]{\includegraphics[width=15cm]{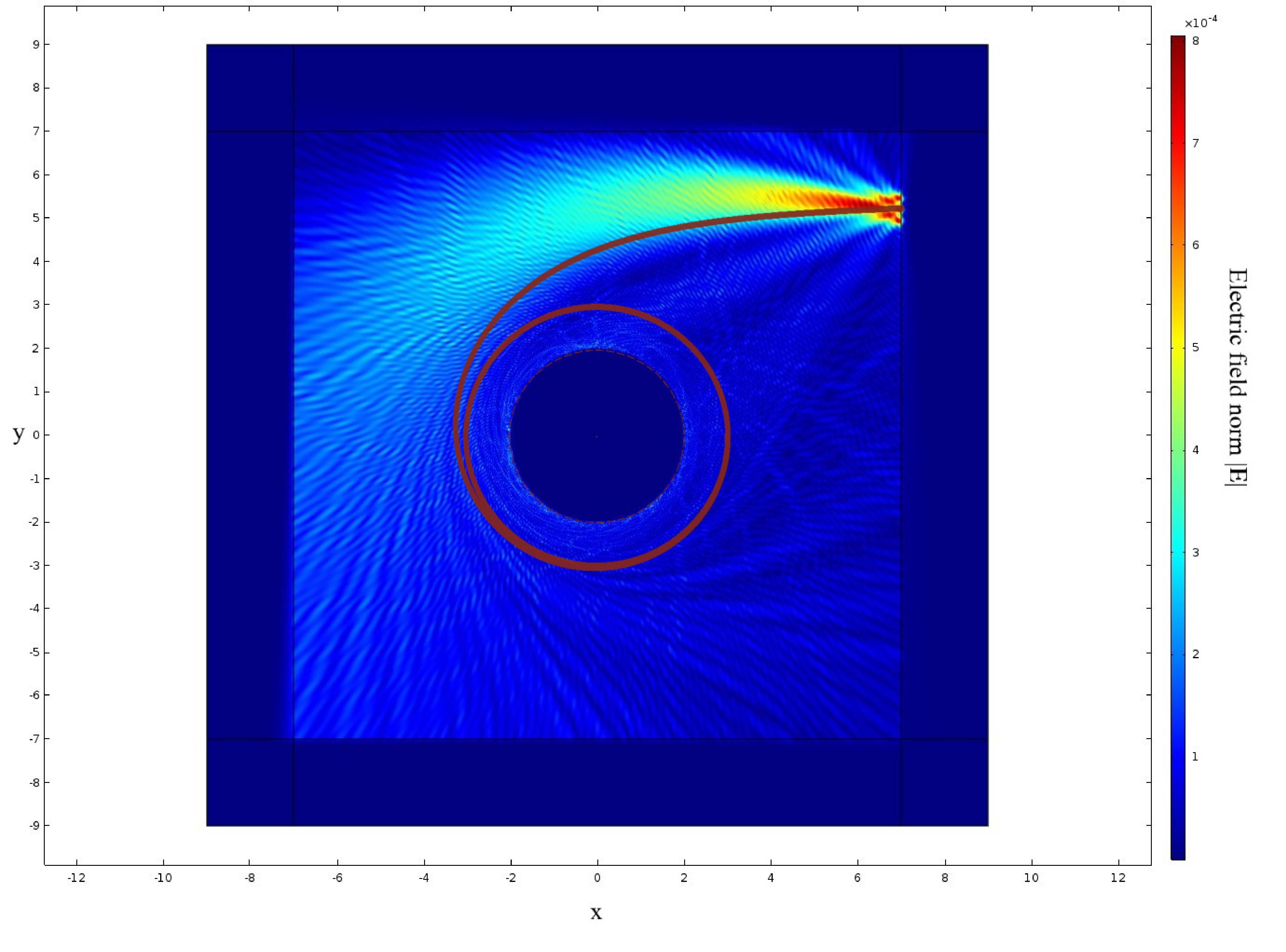}
        \label{RNm1Q1D52final}}
\end{figure}

\begin{figure}
        \centering
    \subfloat[$D=6.2$]{\includegraphics[width=15cm]{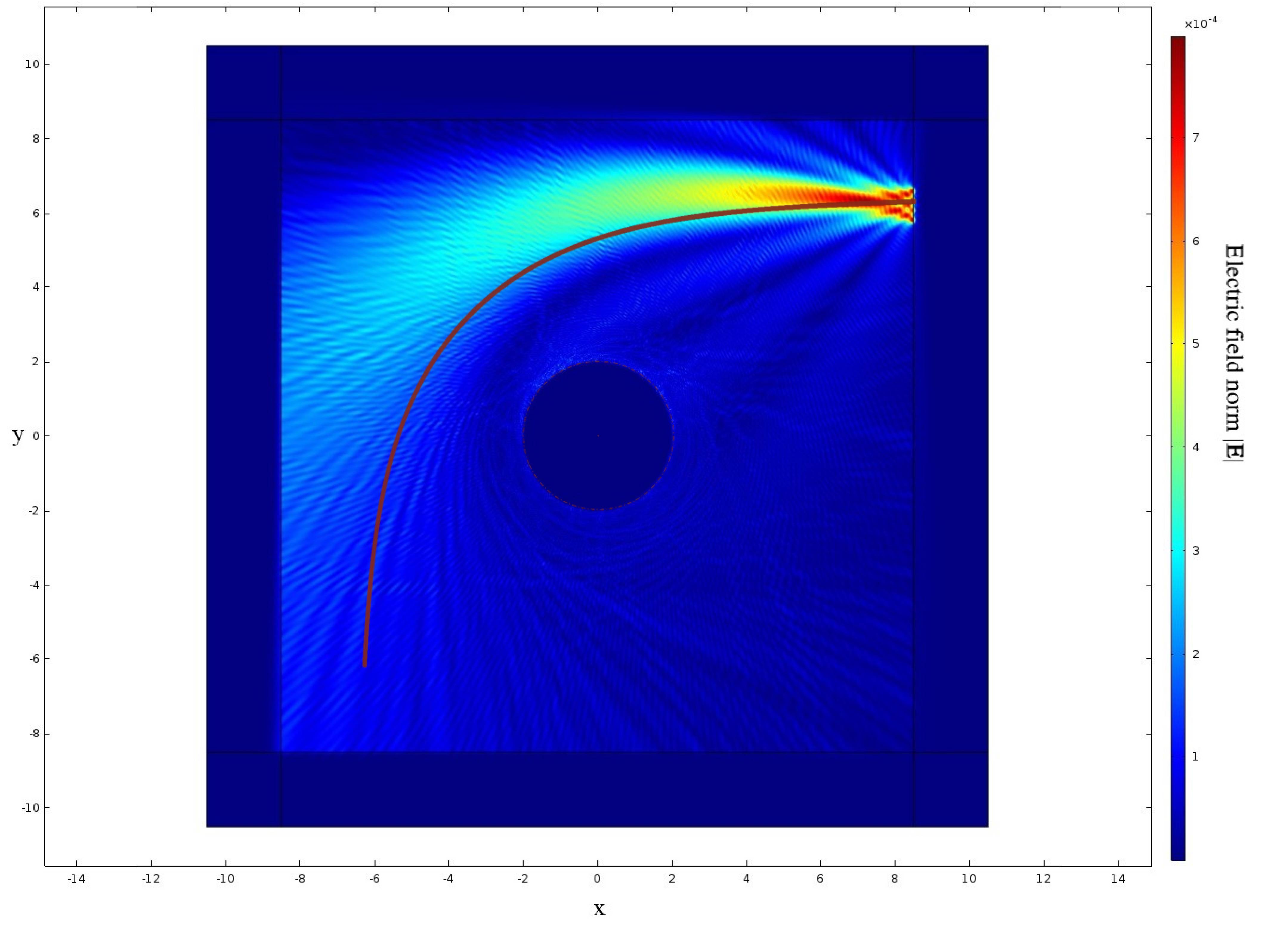}\label{RNm1Q1D62final}}
    \label{rn01} \caption{metamaterial mimicking the Reissner-Nordstr\"{o}m spacetime
        for $Q=0.1$, $m=1$ and different impact parameters, $D$.}
\end{figure}

\begin{figure}[H]
    \centering
    \subfloat[$D=2.97$]{\includegraphics[width=15cm]{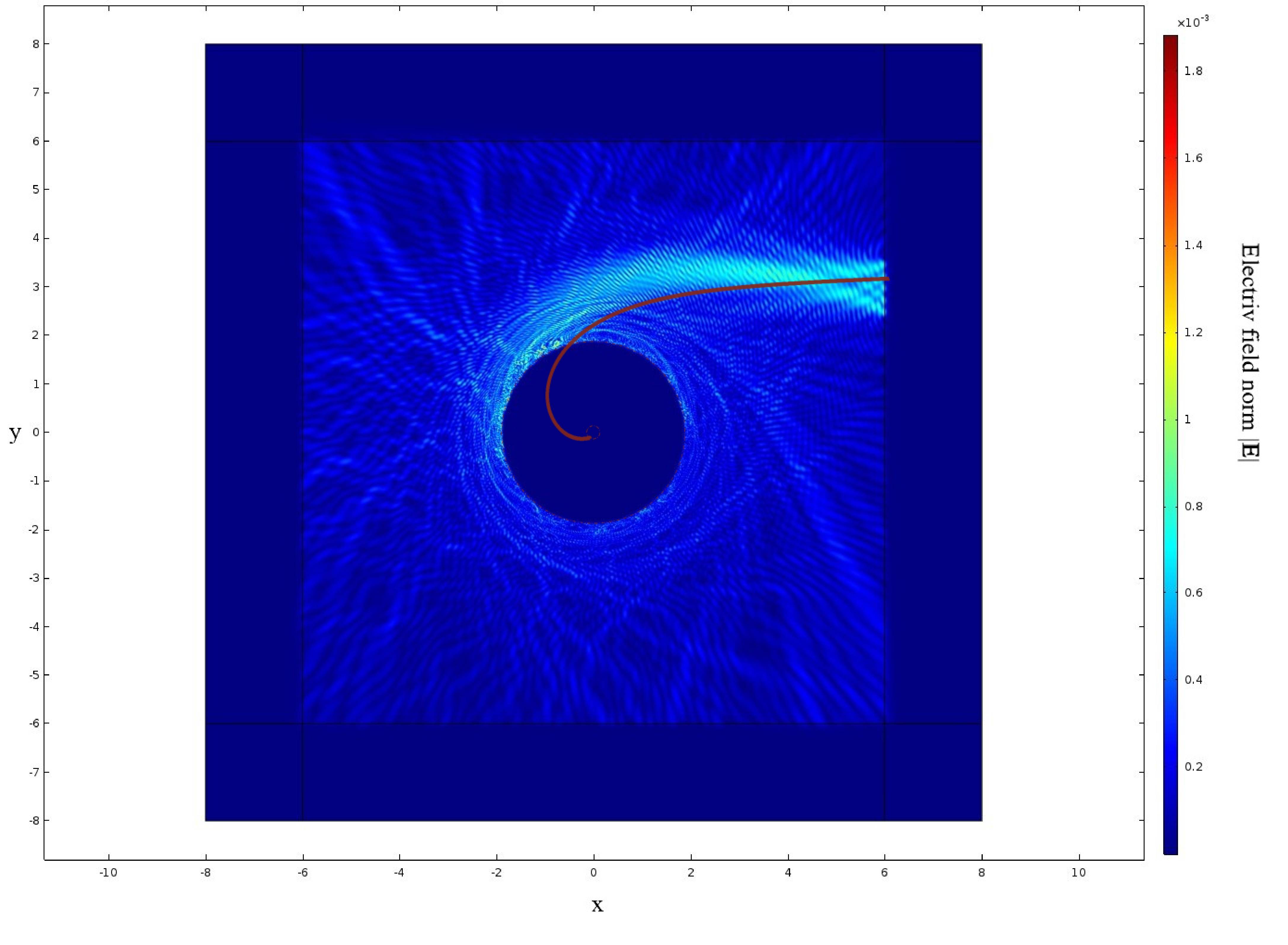}\label{RNm1Q5D297final}}
    \\
    \subfloat[$D=4.97$]{\includegraphics[width=15cm]{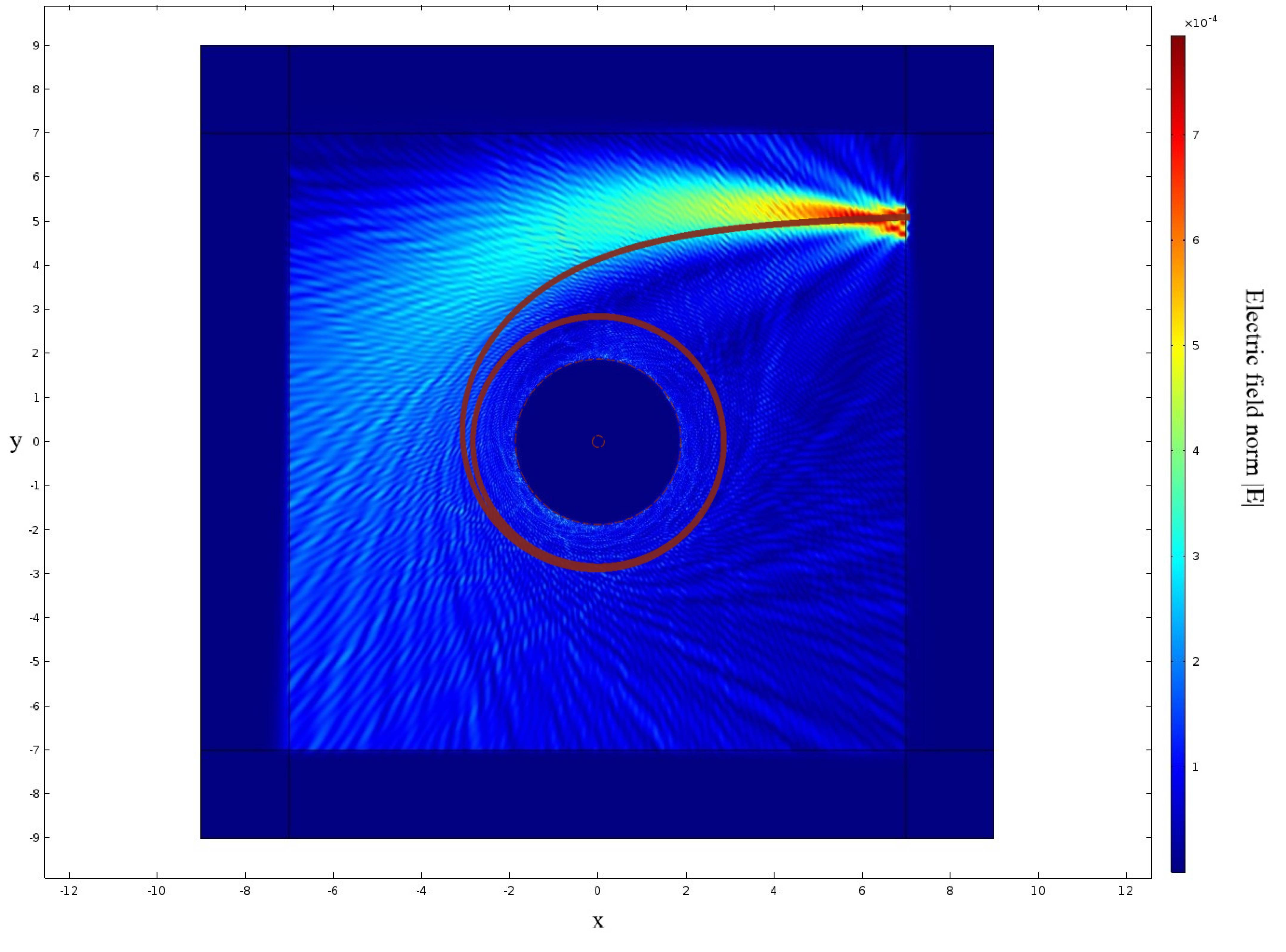}
        \label{RNm1Q5D497final}}
\end{figure}

\begin{figure}
        \centering
    \subfloat[$D=5.97$]{\includegraphics[width=15cm]{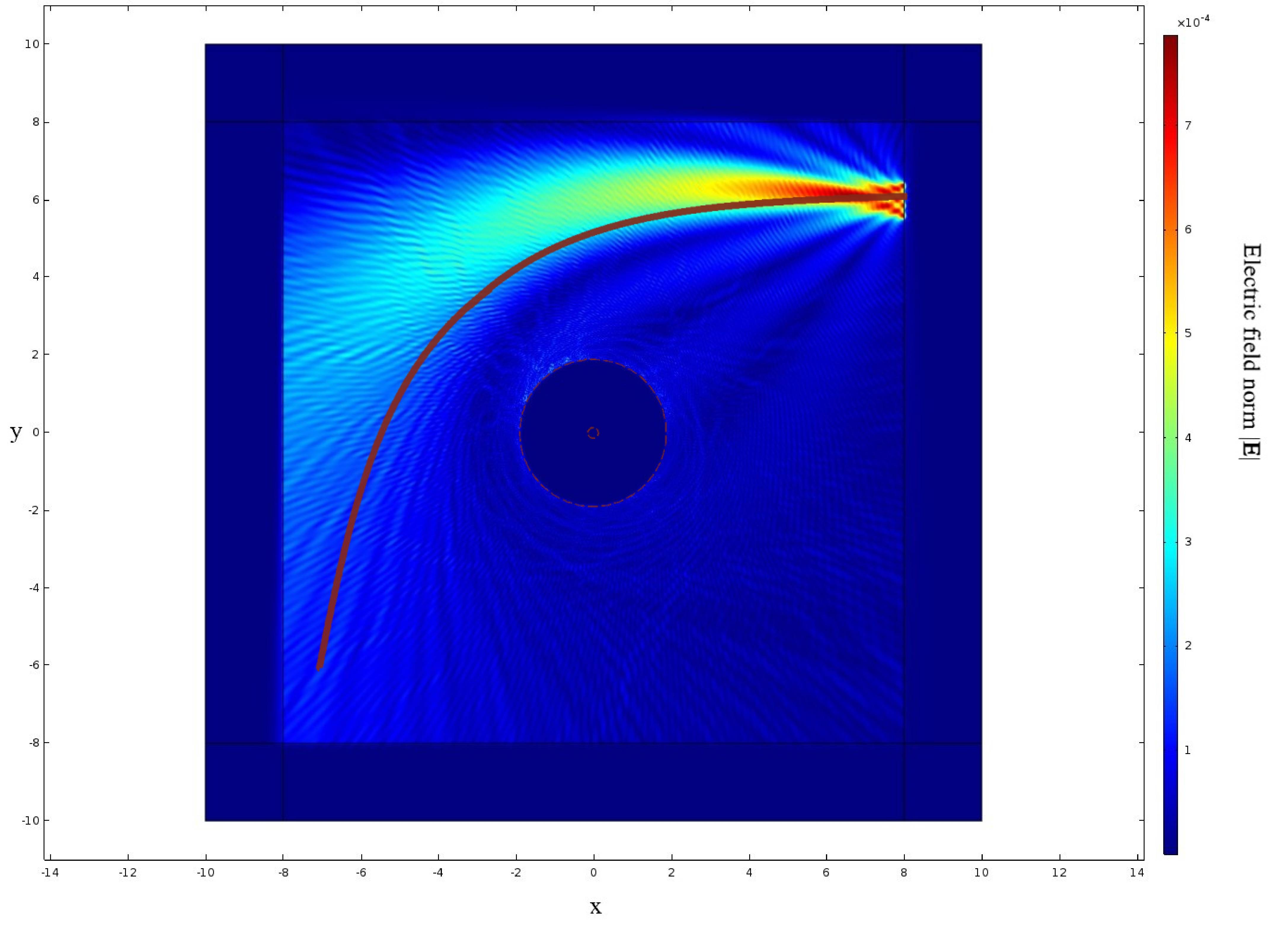}\label{RNm1Q5D597final}}
    \label{rn05}
    \caption{metamaterial mimicking the Reissner-Nordstr\"{o}m spacetime
        for $Q=0.5$, $m=1$ and different impact parameters, $D$.}
\end{figure}

\begin{figure}[H]
    \centering
    \subfloat[$D=2$]{\includegraphics[width=15cm]{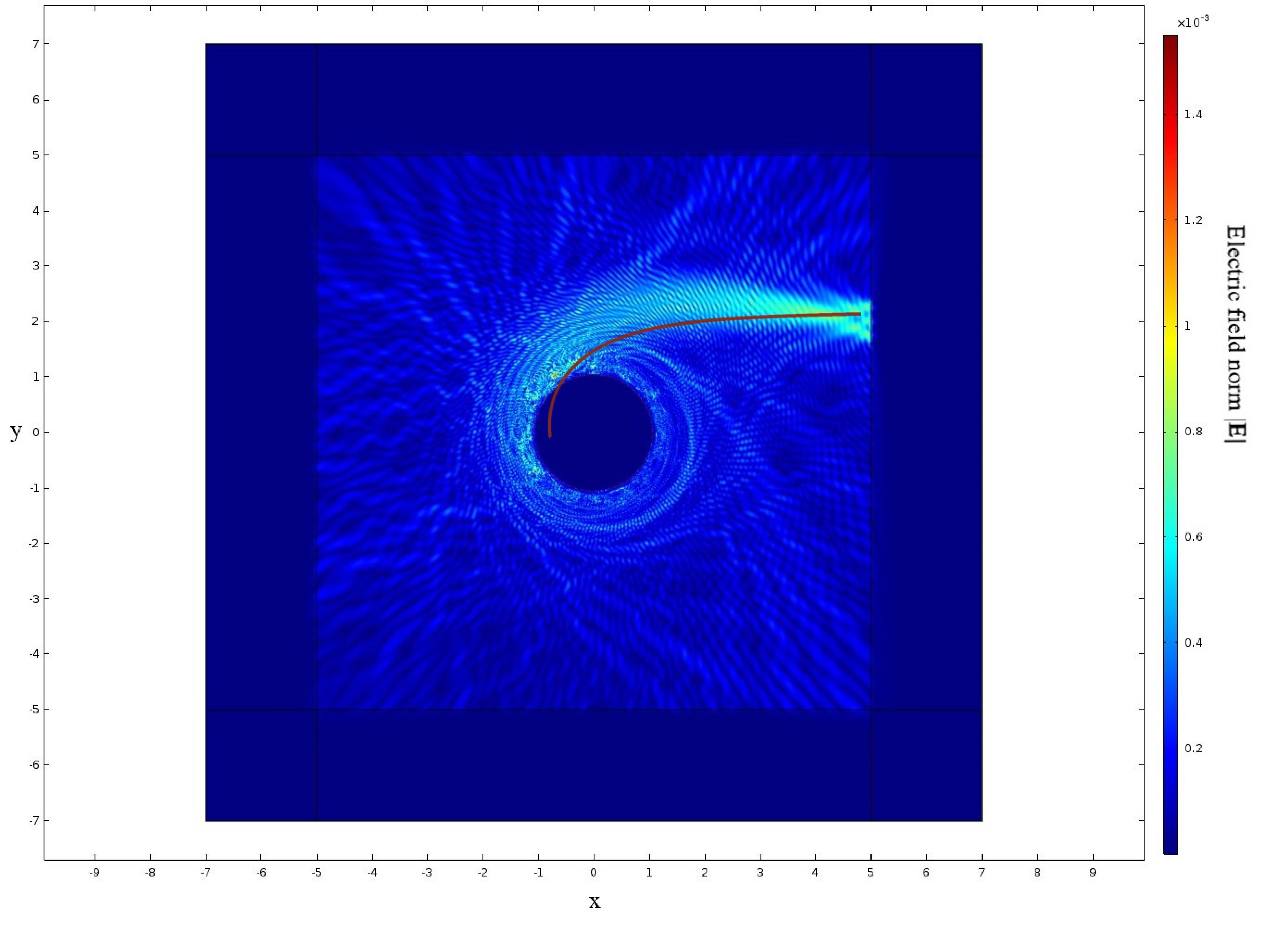}\label{RNm1Q1D2final}}
    \\
    \subfloat[$D=4$]{\includegraphics[width=15cm]{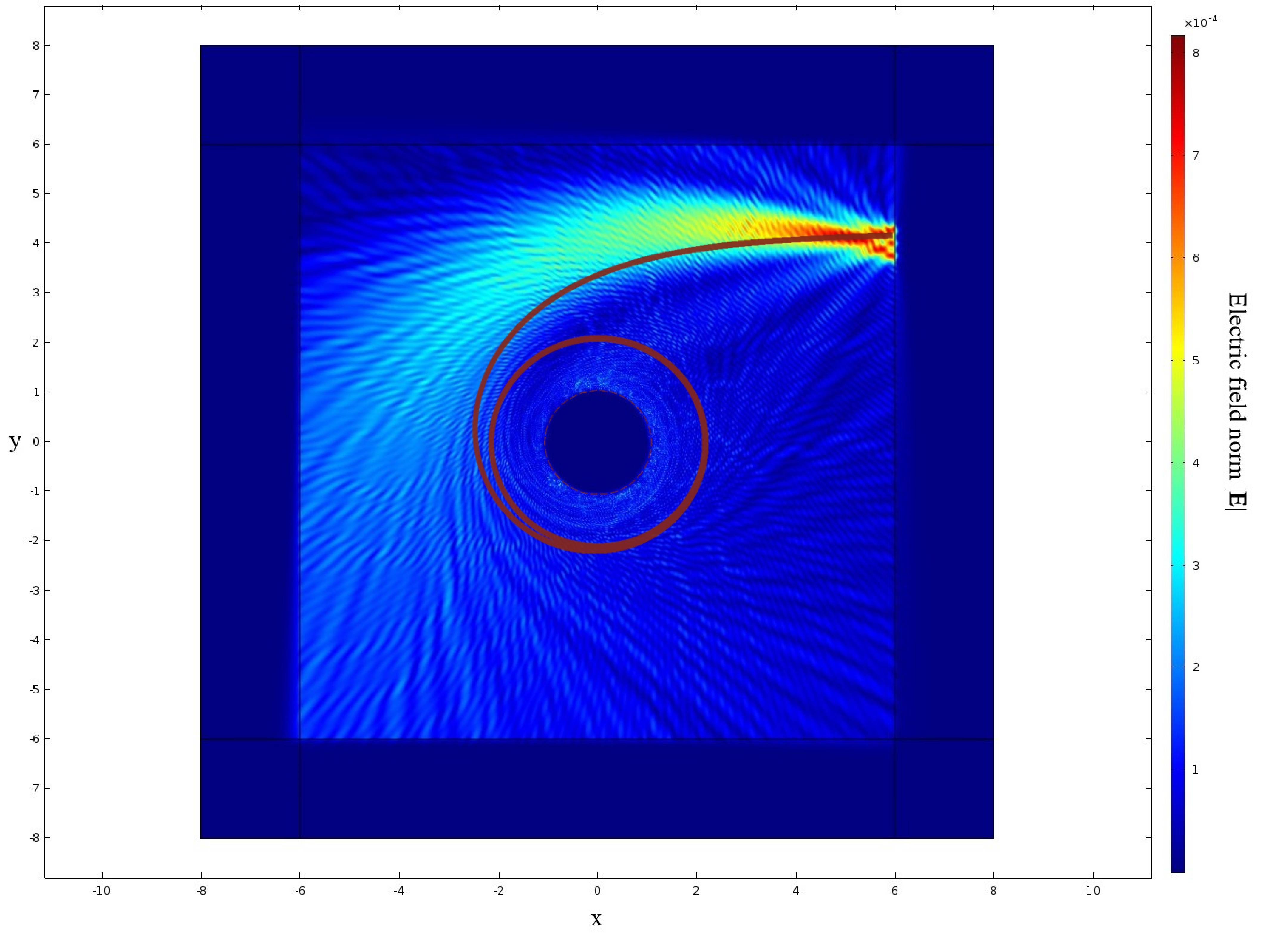}
        \label{RNm1Q1D4final}}
\end{figure}

\begin{figure}
    \centering
    \subfloat[$D=6$]{\includegraphics[width=15cm]{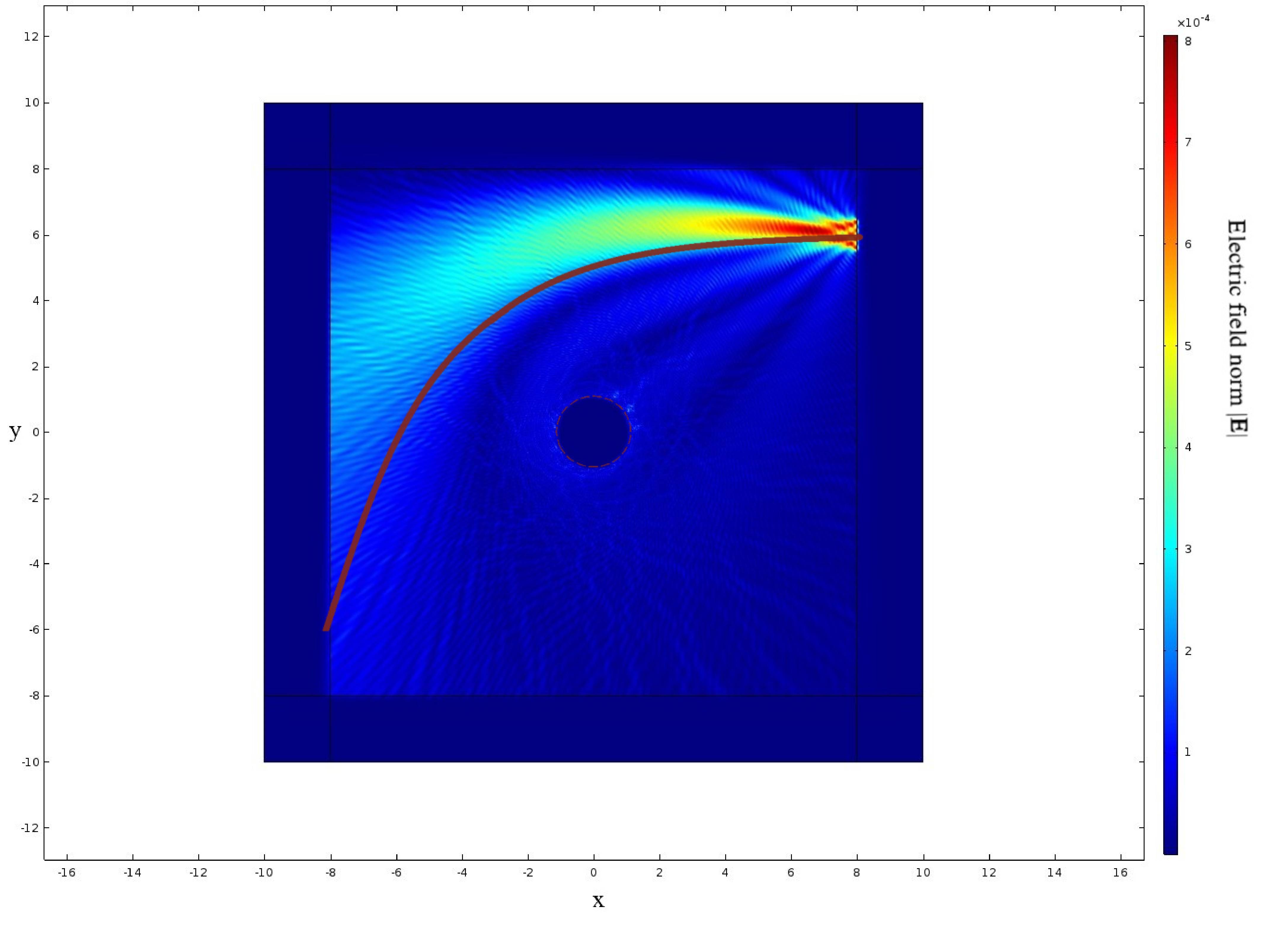}\label{RNm1Q1D6final}}
    \label{rn1}
    \caption{metamaterial mimicking the Reissner-Nordstr\"{o}m spacetime
        for $Q=1$, $m=1$ and different impact parameters, $D$.}
\end{figure}

Taking into account the plotted figures, we find that the ray path
follows the center of the beam. Moreover, the ray trajectories are
consistent with the behavior expected from theory. Speaking more
precisely, the case of $D=D_c$ represents photon sphere
\cite{photonsphere} which is a spherical region of space where the
gravitational effect is strong enough that photons are forced to
travel in an orbit (in the $xy$ plane). This area is patently
obvious in Figs. \ref{schwarzm1D52final}, \ref{RNm1Q1D52final},
\ref{RNm1Q5D497final} and \ref{RNm1Q1D4final}. The photon sphere
is located farther from the center of a black hole than the event
horizon, the radius of photon sphere can be determined as $r_c$ in
Eq. (\ref{eq:dcrcrn}) and some numerical results of $r_c$ can be
found in Table \ref{tabrn}. In addition, for $D<D_c$, we expect
that the photons are drown into the event horizon, it can be seen
in Figs. \ref{schwarzm1D32final}, \ref{RNm1Q1D32final},
\ref{RNm1Q5D297final} and \ref{RNm1Q1D2final}. Moreover, in the
case of $D>D_c$, deflection of ray trajectory can be observed in
Figs. \ref{schwarzm1D62final}, \ref{RNm1Q1D62final},
\ref{RNm1Q5D597final} and \ref{RNm1Q1D6final}. Furthermore, it is
notable that the charge of black hole, $Q$, affects the critical
value of impact parameter, $D_c$, and as a result the radius of
photon sphere, $r_c$ is changed. It can be seen from Table
\ref{tabrn} that when $Q$ increases, the critical impact
parameter, $D_c$, and radius of photon sphere, $r_c$, decrease.

It can be observed from all figures that the beam splits into a
set of rays or sub-beams, one part falls into black hole due to
having $D<D_c$, the other part escapes from black hole because of
$D>D_c$ and another part bends around the photon sphere of the
black hole and interferes with the primary beam.

It is expected from theory that the case of $Q=0$ represents
Schwarzschild spacetime. The radius of photon sphere is determined
$r_c=3m$ for $Q=0$ that is similar to our expectation, also Fig. 1
matches the results reported in Ref. \cite{schwarz3}.

For the Kerr spacetime the medium parameters are given by Eqs.
(\ref{eq:kerrconsz0}) and (\ref{eq:kerrgammaz0}), while the
numerical and simulated results are shown in Table \ref{tabkerr}
and Fig. 5. It is notable that in order to have dimensionless
parameters, they are normalized to $r_s$, as before.

%%%%%%%%%%%%%%%%%%%%%%%%%%%%%%%%%%%%%%%%%%
\begin{table}[H]
    \begin{center}
        \begin{tabular}{|c|c|c|}
            \hline\hline $a$ & $D_{c}$ & $r_c$\\
            \hline\hline $+0.1$ & $2.48$ & $1.38$\\
            \hline $0$ & $2.60$ & $1.5$\\
            \hline
            $-0.1$ & $2.74$ & $1.61$\\
            \hline$-0.5$ & $3.5$ & $2$\\ \hline
        \end{tabular}%
        \\[0pt]
        \caption{Kerr solutions: the numerical results of
            $D_c$ and
            $r_c$
            for different values of $a$ in the case of $m=1$.} \label{tabkerr}
    \end{center}
\end{table}
%%%%%%%%%%%%%%%%%%%%%%%%%%%%%%%%%%%%%%%%%%

As we mentioned before, computational analysis of rotating
spacetime is complicated, and therefore, we just focus on the
critical values of the impact parameter which lead to photon
sphere. In contrast to the Reissner-Nordstr\"{o}m black hole, the
Kerr black hole is not spherical symmetry, but only enjoys axially
symmetry, which has profound consequences for the photon orbits. A
circular orbit can only exist in the equatorial plane
\cite{photonsphere}. Fortunately, both methods of simulations,
numeric solutions to Maxwell equations and the ray trajectory
arisen from the Hamiltonian formalism, are in agreement to each
others. As it is depicted in Table \ref{tabkerr}, by decreasing
the spin parameter ($a$), the critical impact parameter ($D_c$)
and the radius of photon sphere ($r_c$) increase. In addition, in
the case of static case ($a=0$), the results of the Kerr black
hole match to that of driven from Schwarzschild black hole, as
expected.

\section{Concluding Remarks}

Here, we compared geometrical description of gravitational systems
with optical media in flat space with two approaches. We observed
a good correspondence between the two methods.

We considered two supplementary factors of the Schwarzschild black
hole; the electric charge and rotation parameters. More precisely,
we investigated the effects of electric charge and rotation
factors on the trajectory of light with various impact parameters.
We found that the medium parameters given for
Reissner-Nordstr\"{o}m and Kerr spacetime can mimic the ray path
near the black hole in the flat spacetime. Interestingly, some
theoretical phenomena like photon sphere are observed and the
effects of charge, $Q$, and spin parameter, $a$, are investigated.
We showed that the critical impact parameter is a decreasing
function of both the electric charge and spin parameters. This
behavior is expected due to the fact that increasing the electric
charge and rotation lead to decreasing the event horizon radius
(weaken the gravitational effect) in the classical black hole
scenario.

It will be interesting to apply these calculations to the
Kerr–Newman black hole and other interesting nontrivial solutions
of Einstein gravity.

%%%%%%%%%%%%%%%%%%%%%%%%%%%%%%%%%%%%%%%%%%

%%%%%%%%%%%%%%%%%%%%%%%%%%%%%%%%%%%%%%%%%%

%%%%%%%%%%%%%%%%%%%%%%%%%%%%%%%%%%%%%%%%%%
%%%%%%%%%%%%%%%%%%%%%%%%%%%%%%%%%%%%%%%%%%
%%%%%%%%%%%%%%%%%%%%%%%%%%%%%%%%%%%%%%%%%%
\begin{figure}[H]
    \centering
    \subfloat[$a=+0.1, D=2.48$]{\includegraphics[width=15cm]{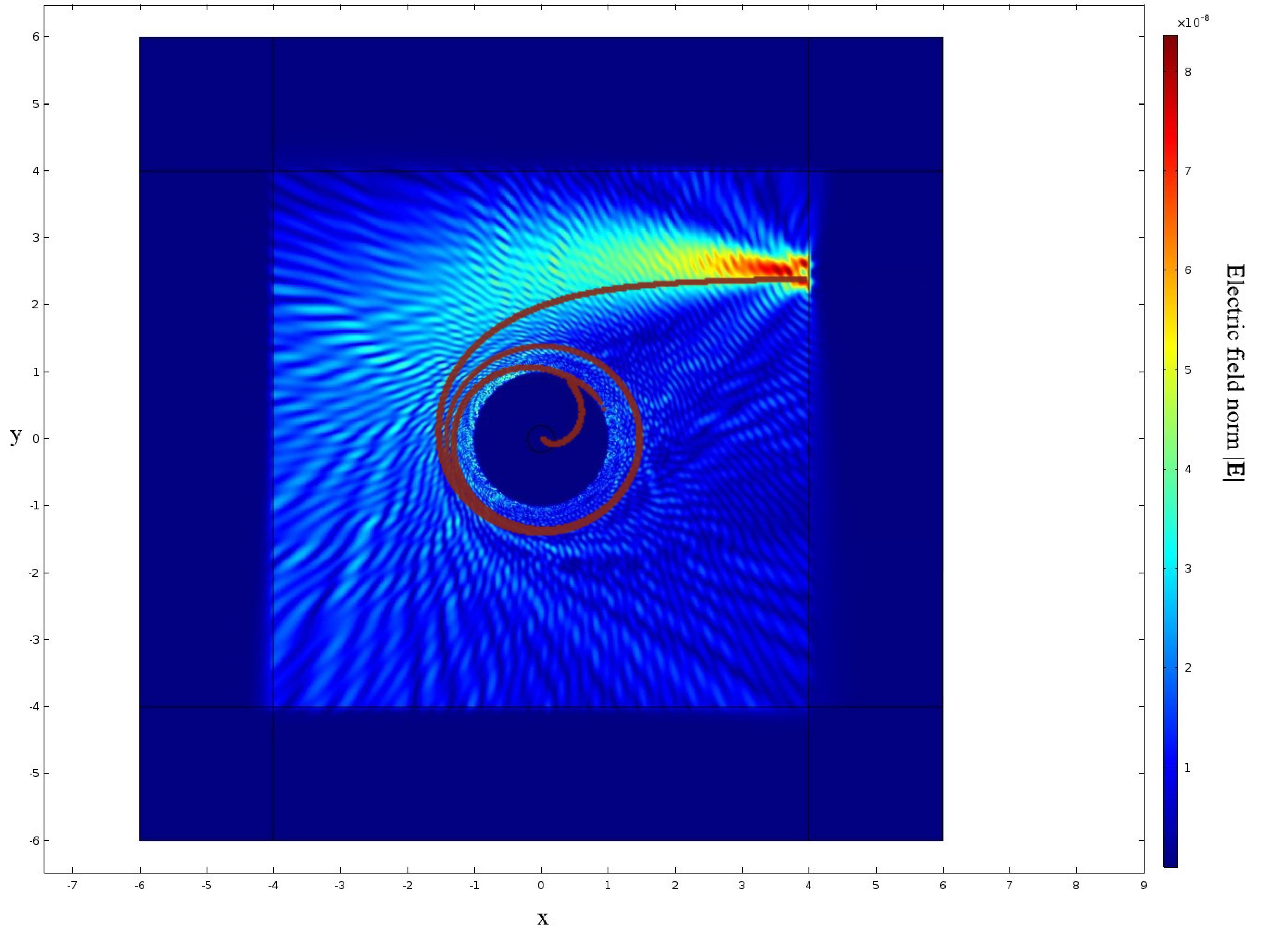}\label{m05a01final}}
    \\
    \subfloat[$a=-0.1, D=2.74$]{\includegraphics[width=15cm]{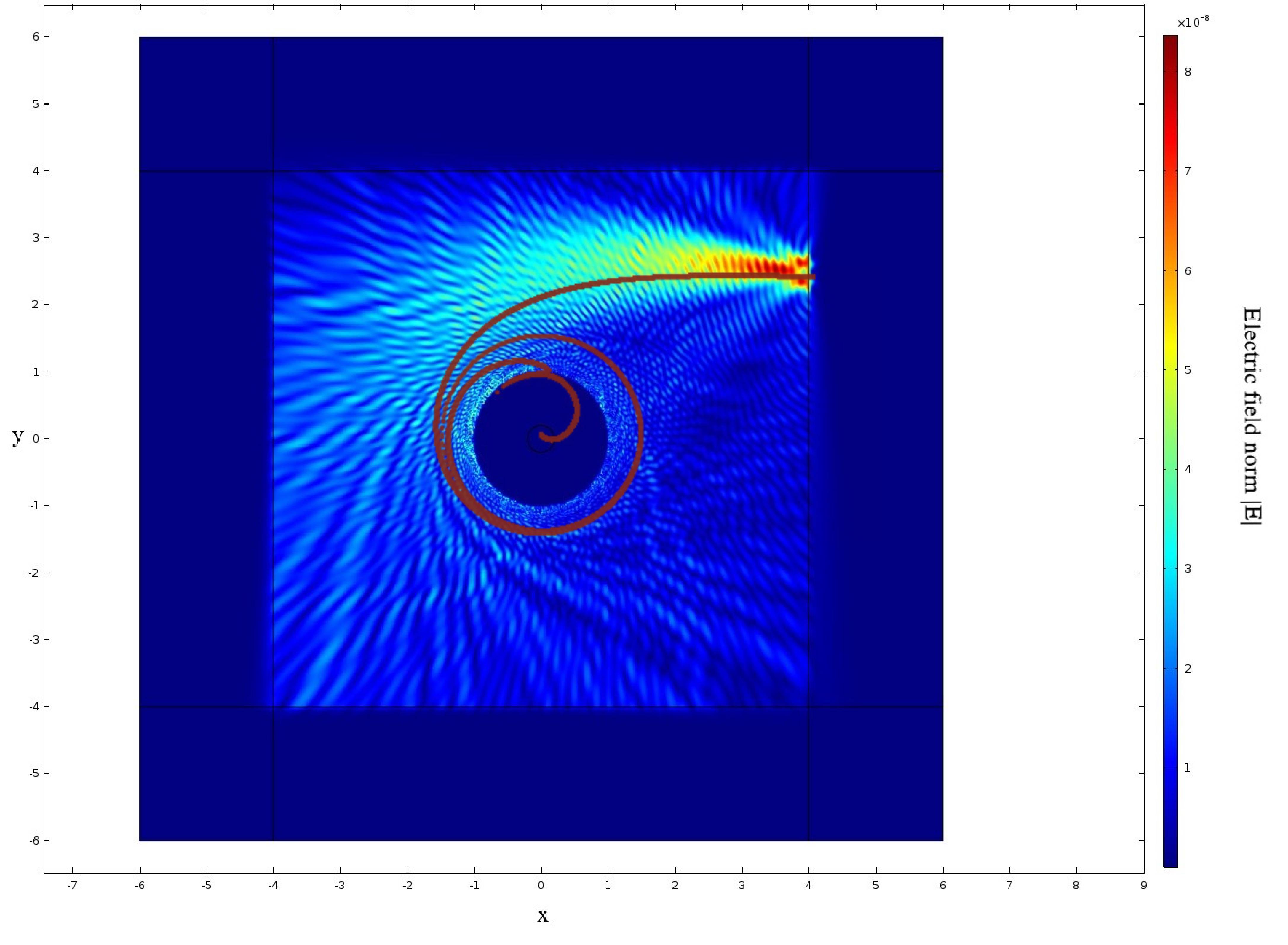}
        \label{m05am01final}}
\end{figure}
\begin{figure}
    \centering
    \subfloat[$a=-0.5, D=3.5$]{\includegraphics[width=15cm]{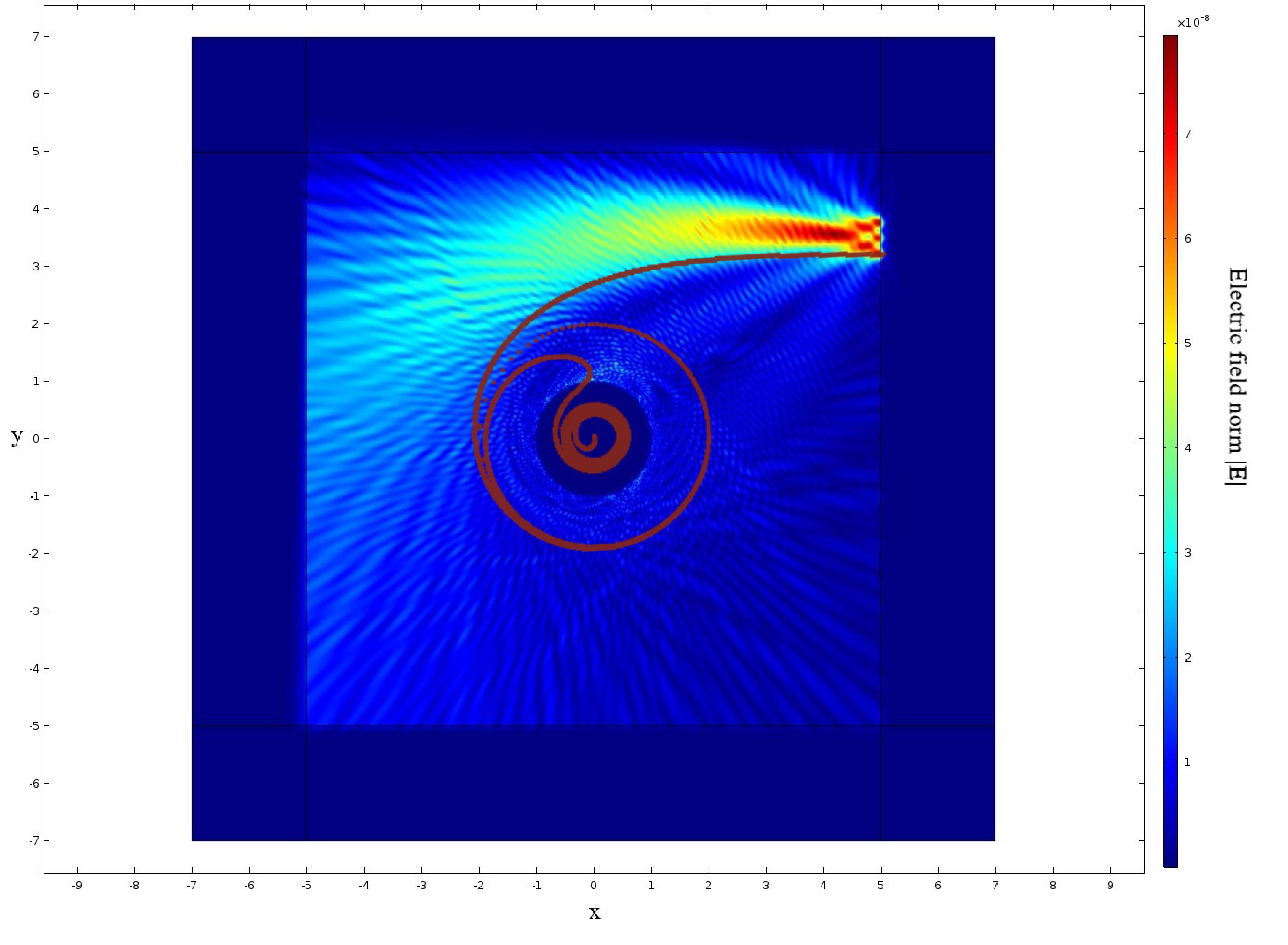}\label{m05am05final}}
    \label{kerr}
    \caption{metamaterial mimicking the Kerr spacetime
        for $m=0.5$ and different impact parameters, $D$, and spin parameters, $a$.}
\end{figure}
%%%%%%%%%%%%%%%%%%%%%%%%%%%%%%%%%%%%%%%%%%

\section{Acknowledgements }

We wish to thank Shiraz University Research Council. This work has
been supported financially by the Research Institute for Astronomy
and Astrophysics of Maragha, Iran.

\end{document}